\def\year{2019}\relax
\documentclass[letterpaper]{article} 
\usepackage{aaai19}  
\usepackage{times}  
\usepackage{helvet}  
\usepackage{courier}  
\usepackage{url}  
\usepackage{graphicx}  
\usepackage[normalem]{ulem}
\usepackage{enumitem,setspace}
\usepackage{amsmath,amssymb,amsthm}
\usepackage{color}
\usepackage[compact]{titlesec}
\titlespacing{\paragraph}{%
  0pt}{
  0.15\baselineskip}{
  0.8em}

\usepackage{pifont}
\usepackage[dvipsnames]{xcolor}
\definecolor{mydarkgreen}{HTML}{008000}
\definecolor{mydarkred}{HTML}{dc143c}
\definecolor{mydarkyellow}{HTML}{ff8c00}

\def\given{\,|\,}
\newcommand\eq{\mbox{\it eq}}

\usepackage{array, makecell}
\usepackage{arydshln, cellspace}
\usepackage{booktabs}
\usepackage{multirow}
\usepackage{makecell}
\newcommand{\citet}[1]{\citeauthor{#1} \shortcite{#1}}

\begin{document}
%
\setlength{\abovedisplayskip}{4pt}
\setlength{\belowdisplayskip}{4pt}

\title{TopicEq: A Joint Topic and Mathematical Equation Model for Scientific Texts}

\author{Michihiro Yasunaga
\,
\quad
John D. Lafferty
\\[1mm]
Yale University\\
\scalebox{0.85}[0.9]{{\tt michihiro.yasunaga@yale.edu}}
}
\maketitle
\begin{abstract}
\vspace{-2mm}
 Scientific documents rely on both mathematics and text to communicate ideas.
 Inspired by the topical correspondence between mathematical equations and word contexts observed in scientific texts,
 we propose a novel topic model that jointly generates mathematical
 equations and their surrounding text (\textit{TopicEq}). Using an extension of the
 correlated topic model, the context is generated from a mixture of
 latent topics, and the equation is generated by an RNN
 that depends on the latent topic activations. To experiment with this model,
 we create a corpus of 400K equation-context pairs extracted from a range of scientific articles from arXiv,
 and fit the model using a variational autoencoder approach.
 Experimental results show that this joint model significantly outperforms existing topic models and equation models for scientific texts.
 Moreover, we qualitatively show that the model effectively captures the relationship between topics and mathematics, enabling novel applications such as topic-aware equation  generation, equation topic inference, and topic-aware alignment of mathematical symbols and words.
 \vspace{-2mm}
\end{abstract}

\section{Introduction}
\label{sec:intro}

Technical scientific articles, such as those from physics and computer science, rely on both mathematics and text to communicate ideas.
Most existing work in natural language processing (NLP) and machine learning  studies these two components separately.
For instance, text-based
topic models have been used widely on scientific articles to uncover their semantic structure \cite{blei2003latent,blei2006dynamic,newman2010visualizing}.
For mathematics, recent work \cite{lan2015mathematical,zanibbi2016multi,deng2017image} has studied methods to model and generate mathematical equations, for example using RNNs.
However, ultimately these two components should be processed together in a seamless manner.
Algorithms for automated understanding of scientific documents
should extract the information encoded by not only words but also mathematical equations.
At the same time, equations should ideally be modeled with the help of the surrounding text,
as the meaning of an equation depends not only on its
constituent symbols and syntax, but also on the context in which it appears
\cite{wang2015wikimirs,krstovski2018eq}.

\begin{figure}[!t]
    \vspace{-1.5mm}
    \hspace{-1.5mm}
    \includegraphics[width=0.48\textwidth]{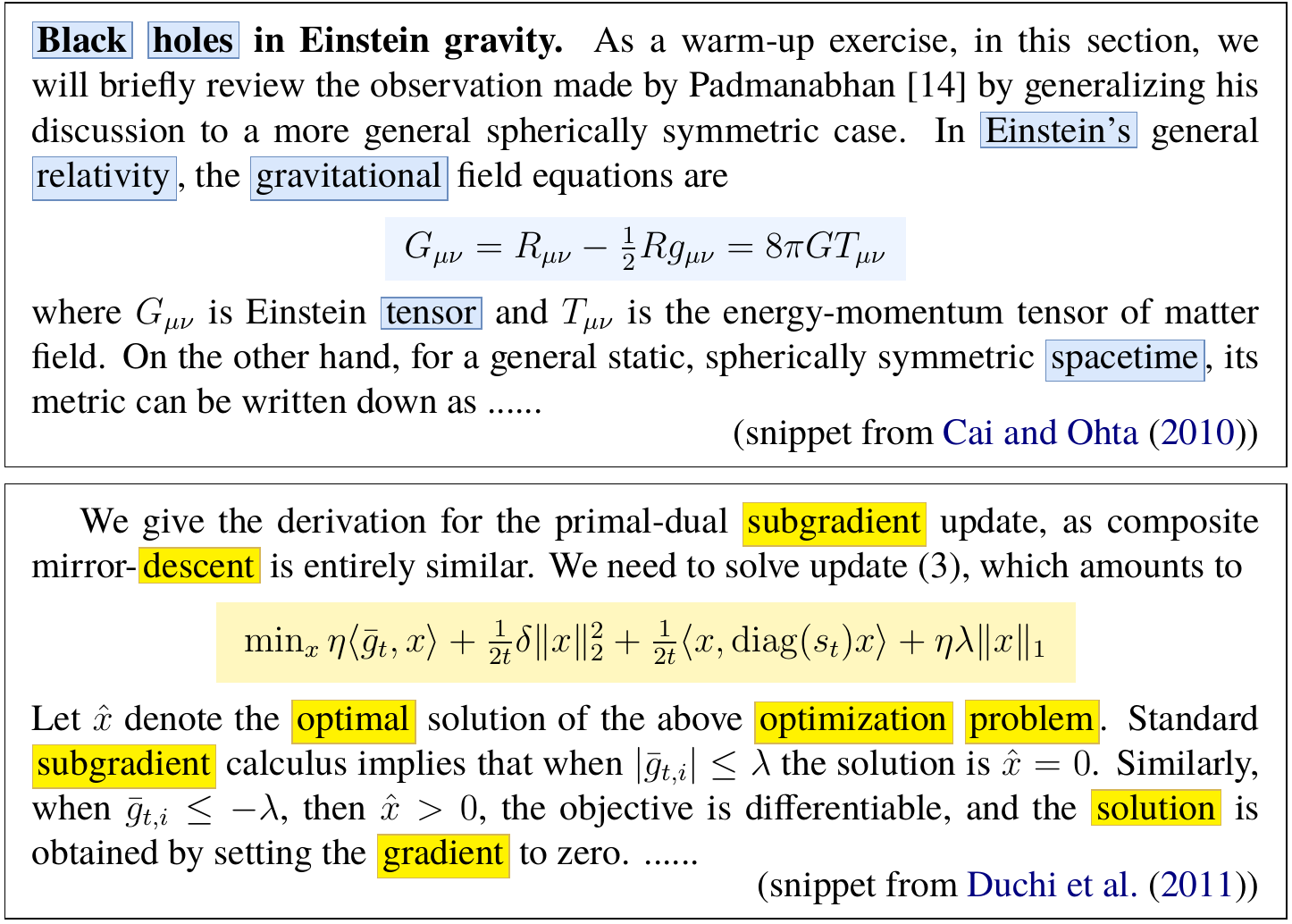}
    \vspace{-7mm}
    \caption{The words in a given technical context often
      characterize the distinctive types of equations used, and  vice versa.
    \textbf{Top} topic: Relativity; \textbf{bottom} topic: Optimization.
    }\vspace{-4mm}
\label{fig:context_exm}
\end{figure}

To this end, this paper proposes
a topic-equation model
that \textit{jointly} generates equations and their surrounding text in scientific documents (\textit{TopicEq}),
and demonstrates that the model can effectively achieve the aforementioned two goals.
The intuition
behind the model is illustrated in the sample passages in
Figure~\ref{fig:context_exm}, which shows how the topic of the word context is often indicative of the distinctive types of equations used, and vice versa.
For instance, equations appearing in the topic of relativity (with context words like ``back hole'', ``Einstein'')
tend to involve a series of tensors like $G_{\mu\nu}$ and $T_{\mu\nu}$, while equations used in the
topic of optimization (context words ``gradient'', ``optimal'') may use norms,
the $\min$
operator, and often their combinations.
Ideally, the strings of mathematical symbols in the equations should aid the training of topic models, and the context words should aid the modeling and understanding of the equations.

Our model formalizes this intuition for
scientific texts by generating each equation and its context passage
using a shared latent topic.  Specifically, we apply a
topic model to the context passage, and use the same latent topic proportion vector in a recurrent neural network
(RNN) to generate the equation as a sequence of symbols.
To develop and experiment with this model, we construct a
large corpus of context-equation pairs, extracted from the \LaTeX\ source of
arXiv articles across a range of scientific domains (\textit{ContextEq-400K}). We fit the model on this corpus
using approximate inference based on a variational autoencoder
approach.

Our evaluation shows that this joint model significantly outperforms alternative topic models and RNN equation models for scientific texts. We further show that the model enables novel applications that bridge topics and mathematical equations.
Concretely,
the paper makes the following contributions.

\begin{itemize}[topsep=0.5mm]
\setlength{\itemsep}{-0mm}
    \item
    The first study of jointly modeling topics and mathematics in scientific texts.

    \item Better topic models for scientific texts: Joint training with the RNN equation model boosts the quality of topic modeling. This greatly outperforms the topic model that includes equations simply as bags of tokens, suggesting that equations' syntax-level information captured by the RNN is useful for topic modeling.

    \item Better equation models: Joint topic modeling
    provides the narrative context for equation prediction, and improves the quality/grammaticality of the RNN equation model.

    \item
    Our model successfully captures the relationship between mathematical equations and topics (words), enabling interpretable handling of equations.
    For instance, we illustrate that the model enables
    topic-aware equation generation and equation topic inference. We also present a variant of this model that learns topic-aware associations between mathematical symbols and words.

    \item The model is unsupervised, and enables the aforementioned tasks and applications without manual labels.
\end{itemize}

\section{Related Work}
\label{sec:related_work}
Our work is connected to a wide range of recent research, from topic models to mathematical equation processing.

\paragraph{Topic models.}
Topic models provide a powerful tool to extract the semantic structure of texts in the form of the latent topics---usually multinomial distributions over words.
Starting from LDA \cite{blei2003latent}, topic models have been studied extensively
\cite{teh2005sharing,blei2006dynamic,blei2007correlated,hall2008studying}, especially for scientific articles.
However, while mathematical equations play an essential role in scientific documents, topic models capable of processing equations besides word texts are yet to be studied.
This work shows that incorporating joint modeling of equations via an RNN boosts the performance of topic modeling for scientific texts.

Recent work
\cite{cao2015novel,larochelle2012neural} has proposed neural topic models, leveraging the
flexibility and representation power of neural networks.
In particular, \cite{miao2016neural,miao2017_disc,srivastava2017autoencoding} employ neural variational inference to
train topic models; we will apply their technique to fit our model.

\paragraph{Language models \& equation models.}

Language modeling aims to learn a probability distribution over a sequence of words.
It is a fundamental task in NLP, with a plethora of applications including text generation.
RNN-based language models are shown effective for sequences with
long-term dependencies
\cite{mikolov2010recurrent,jozefowicz2016exploring}.

Similar to language models, equation models are useful for various tasks involving equation generation, such as semantic parsing \cite{roy2016equation} and handwriting / optical character recognition \cite{deng2017image}.
The use of RNNs to model \LaTeX\ was illustrated by \cite{Karpathyblog} for an algebraic geometry text.
This work also employs an RNN to model each equation
as a sequence of \LaTeX\ tokens (or ``symbols,'' interchangeably).

\paragraph{Neural topic-language models.}
Our model architecture is motivated by
joint topic-language models. Such models typically extract latent topics of a given document via a topic
model, and utilize the topic knowledge to improve an RNN language model.
\citet{mikolov2012context} incorporate the topic vector of a
pre-trained LDA model into an RNN language model; recent work
\cite{dieng2016topicrnn,lau2017topically,wang2017topic}
trains neural topic and language models jointly, as we will do here.

Key distinctions can be made between our work and these models.
First, while previous work uses topic models to improve language modeling on the same word text, our task models two different modalities: word text and equations. In this sense, our work is related to
\cite{blei2003modeling}, which extends LDA to model image-text pairs. Moreover, taking advantage of these two modalities, we also present a variant of the TopicEq model that learns topic-aware association between mathematical symbols and words.

The second difference lies in the RNN equation model we propose.
While
\cite{dieng2016topicrnn,ahn2016neural,lau2017topically} integrate the
topic knowledge into either the output layer of the LSTM or the word
predictions of the language model, we embed the topic proportion vector
inside the LSTM, to enable the topic knowledge to have
deeper influence on equation generation.
Experimental results show that this method of incorporating topic information
is more effective than the existing methods for improving the quality of
equation modeling.

\paragraph{Mathematical equation processing.}
Some
work has processed equations as bags of math symbols to extract their features for searching
\cite{sojka2011indexing} and clustering \cite{lan2015mathematical}.
\citet{zanibbi2016multi} introduce tree-based representations for equations for mathematical information retrieval tasks.
Most recently, \citet{deng2017image} propose
RNN-based models to generate equations.
We will show that RNN-based equation processing can \textcolor{black}{capture syntactic features of equations}, and provides more effective help for topic modeling than bag of token-based equation processing does.

Finally, our work of modeling equations with contexts is  related to \cite{krstovski2018eq}, which fits equation embeddings using surrounding words.
While they limit the equation domains (i.e., ML, AI),
this work aims to uncover topics for texts and equations from a range of scientific domains.
This work also models each equation itself as a sequence of symbols, which is not studied in their work.

\section{The TopicEq Model}

\begin{figure}[!t]
    \vspace{-0mm}\hspace{-1mm}
    \centering
    \includegraphics[width=0.37\textwidth]{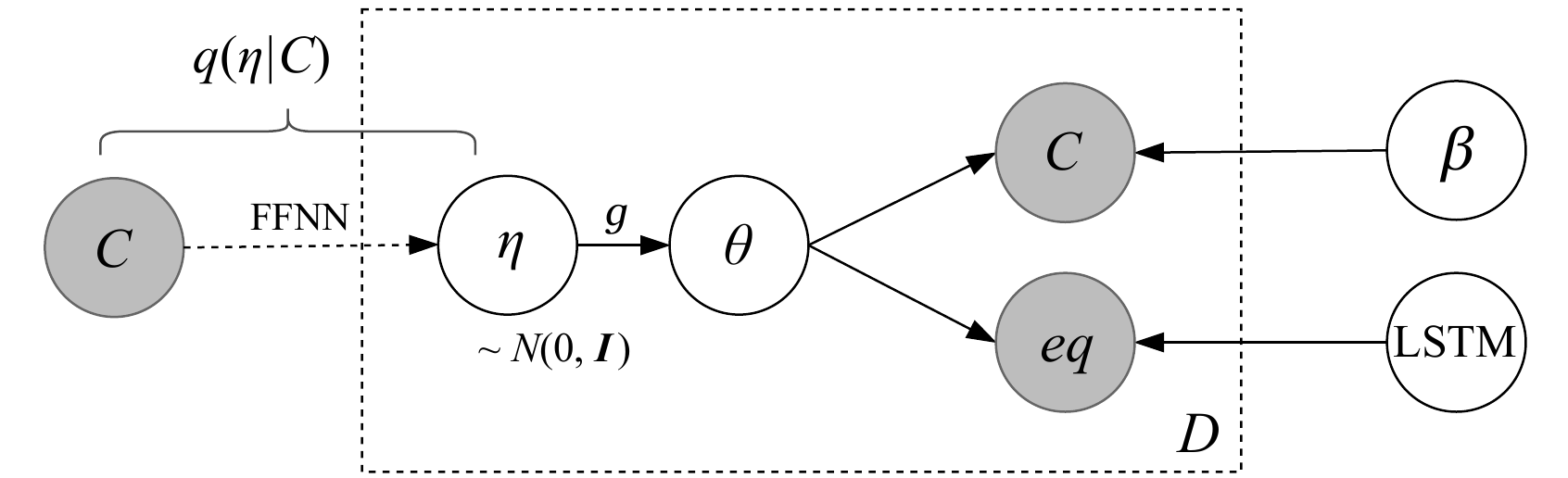}\vspace{-2mm}
    \caption{Graphical structure underlying the TopicEq model.}
\label{fig:topic_eq_model}
\vspace{-4mm}
\end{figure}

Our starting point is the correlated topic model
\cite{blei2007correlated}, which models the topic proportion
vector through a latent Gaussian vector. We extend this model
to the setting where each ``document''\vspace{-1pt} consists of a
displayed equation $\eq$ and its surrounding text $C=
\{w_n^{(c)}\}_{n=1}^N$, which we call the equation's \textit{context}.
Our joint model assumes that each equation and its context are generated from the same latent topic vector $\theta$; see Figure \ref{fig:topic_eq_model}.
Concretely, the generative process for a given
$D = (C,\eq)$ is
\begin{align}
    \label{eq:theta_prior}
    \eta \sim \mathcal{N} (0,I), ~~~ \theta = g(\eta)
\end{align} \vspace{-5.5mm}
\begin{align}
    & w_n^{(c)} \mid \theta \sim \text{Mult}(\theta^T\beta) \label{eq:wn_theta}\\
    & \eq \mid \theta \sim \text{LSTM}(\theta)  \label{eq:lstm_theta}
\end{align}
where $g (\eta) \!=\! \mathrm{softmax}( W_g\eta + b_g)$. Note that this is equivalent to placing a logistic normal distribution
on $\theta$ where the latent Gaussian has mean $b_g$ and covariance
$W_g W_g^T$. The parameters $W_g,
b_g$, the topics $\beta$, and the weights in the LSTM are to be estimated from data.
 Expressing the model as shown in Figure~\ref{fig:topic_eq_model} emphasizes the
connection with neural topic models such as
\cite{miao2017_disc}; we will apply their model training technique.

Both the words and the equation are generated in a way that depends on
the topic proportion vector $\theta$.
The topics $\beta^T = (\beta_1,
\dots, \beta_K )$\vspace{-2pt} are distributions over a word vocabulary with $V$ words;
the context words $w_n^{(c)}$ are then drawn from the mixture $\theta^T \beta$, similar to \cite{wang2017topic}.
We employ an RNN
to generate $\eq$ as a sequence of mathematical tokens, where the
vocabulary is extracted from the set of \LaTeX\ tokens.
Specifically,  to generate an equation conditioned on the latent topic proportion
vector $\theta$ (equivalently $\eta$),
we consider a {\it Topic-Embedded LSTM} (TE-LSTM), an extension of the LSTM \cite{Hochreiter:1997:LSM:1246443.1246450}
where the $t$-th update is\vspace{-1mm}
\begin{align*}
    i_t &= \sigma \left (W_i [x_t ; h_{t-1}; \theta] + b_i \right )\\[-0pt]
    f_t &= \sigma \left (W_f [x_t ; h_{t-1}; \theta] + b_f \right )\\[-0pt]
    \tilde{c}_t &= \tanh \left (W_c [x_t ; h_{t-1}; \theta] + b_c \right )\\[-0pt]
    o_t &= \sigma \left (W_o [x_t ; h_{t-1}; \theta] + b_o \right )
\end{align*}\vspace{-5.7mm}
\begin{align*}
    \scalebox{0.95}[1]{$c_t = f_t \odot c_{t-1} + i_t \odot \tilde{c}_t, ~~~~
    h_t = o_t \odot \tanh \left ( c_t \right ).$}
\end{align*}
Here $[x_t; h_{t-1}; \theta ]$ denotes the concatenation of the current input, previous state and topic proportion vector; $\sigma$ is the sigmoid function and $\odot$ denotes
the Hadamard product.
The probability of the next token in the equation is $p(y_t \,|\, y_{1:t-1})$ $= \mathrm{softmax}$ $(W_y h_t+b_y)$.
Thus, the TE-LSTM embeds $\theta$ inside the LSTM cell to reflect
the topic knowledge for equation generation.
As a joint topic-equation model, it is similar to the
topic-language model of \cite{wang2017topic}.

Writing the equation as a sequence of tokens
$\eq = y_{1:T}$, the training objective is the marginal likelihood of $C$ and $\eq$\\[-5mm]
\begin{align}
	p(C, y_{1:T}) = \int_\eta p(\eta) p(C |\eta)
	\prod_{t=1}^{T}  p(y_t \hspace{0.5pt}|\hspace{0.5pt} y_{1:t-1}, \eta)  d\eta  
\end{align}
~\\[-2.5mm]
Since its direct optimization is intractable, we employ variational inference \cite{jordan1999introduction}.
Denoting the variational distribution by $q(\eta)$, we
maximize the variational lower bound (ELBO) for the log-likelihood, $\log p(C, y_{1:T})$:\\
~\\[-7mm]
\begin{align}
	\mathcal{L} &=   \mathbb{E}_{q(\eta)} \bigl[ \log p( C | \eta ) \bigr]- D_{\text{KL}}\bigl( q(\eta) \,\|\, p( \eta) \bigr)  ~~~ \nonumber\\
	& \hspace{8mm}+ \mathbb{E}_{q(\eta)} \left[ \sum_{t=1}^{T}\log p(y_t \,|\, y_{1:t-1}, \eta) \right]
    \label{eq:elbo}
\end{align}
~\\[-2.5mm]
Following recent approaches to neural topic-language models \cite{miao2017_disc,dieng2016topicrnn,wang2017topic},
we compute $q(\eta)$ as a function of the context $C$
using the variational autoencoder technique \cite{kingma2013auto}.
Specifically, we use a feed-forward neural network (FFNN) as an
inference network to
parameterize the mean and variance vectors of the (diagonal) Gaussian variational
distribution $q(\eta \given C)$. We then
use samples from $q$
to optimize Eq \ref{eq:elbo}.
The parameters of the inference network, the topic model, and the equation model are jointly trained by stochastic gradient descent.

We also include a topic diversity regularization term
to Eq \ref{eq:elbo}, following \cite{xie2015diversifying}.
We observed that this technique prevents learning generic, redundant topics.

\section{Experiments}
We study the performance of the proposed model on a corpus of
context-equation pairs constructed from arXiv articles.
We quantitatively show that our joint topic-equation model provides superior fits than alternative topic models and equation models.
We further demonstrate its efficacy through qualitative analyses and novel applications, such as equation generation and equation topic inference.

\subsection{Dataset Construction (\textit{ContextEq-400K})}
\label{sec:dataset}
To obtain a dataset of context-equation pairs, we used scientific articles
published on arXiv.org.  We sampled 100k articles from all
domains in the past 5 years, and split them into train, validation
and test sets (80\%, 10\%, 10\%).  For each article, we parsed its
\LaTeX\ source and extracted single-line display equations that have
five consecutive sentences both before and after the
equation, which are used to define the word context.  Following
\cite{deng2017image}, we further tokenized each equation into a
sequence of \LaTeX\ tokens (e.g., \verb|\sigma|, \verb|^|, \verb|{|,
\verb|2|, \verb|}|) and kept those of length 20--150, yielding the final corpus
of 400K equation-context pairs.
An equation has 63 tokens on average. The context size of 10 sentences is similar to the document size used in recent work of topic-language models
\cite{dieng2016topicrnn,wang2017topic}.

\subsection{Experimental Setup}

We fit the TopicEq model end-to-end on the train set and evaluate its performance on the test set.

\paragraph{Preprocessing.}
For the topic modeling of context passages, we first removed all the inline math expressions in the text. We then followed the preprocessing steps in \cite{wang2017topic} to tokenize and lowercase all words, exclude stopwords
and words appearing in fewer than 100 documents; this resulted in a vocabulary size of 8,660.
For equations, we use the 1,000 most frequent \LaTeX\ tokens as our vocabulary.

\begin{table}[t!]
\renewcommand{\arraystretch}{1.05}
\addtolength{\tabcolsep}{0pt}
\hspace{-2mm}
\centering
\scalebox{0.90}{
\begin{tabular}{l||cl}
\Xhline{2.5\arrayrulewidth}
     \multicolumn{1}{c||}{\textbf{Topic Model} \vrule width 0pt height 11.5pt depth 0pt} & ~\textbf{50}~ & ~\textbf{100}~  \scalebox{0.7}[0.75]{(\# Topics)\!\!\!\!}\\\Xhline{2.5\arrayrulewidth}
 \!LDA (context only)~\vrule width 0pt height 12.5pt depth 0pt &  .085 & .083 \\
 \!Ours (context only) \vrule width 0pt height 0pt depth 0pt & .085 & .084  \\
 \!Ours (context + Eq BOW)~\vrule width 0pt & .087   &  .086 \\
 \!Ours (context + Eq LSTM) \vrule width 0pt & {\bf .097} & {\bf .094} \\
 \!Ours (context + Eq LSTM shuffled)\! \vrule width 0pt height 0pt depth 5pt & .086 & .085  \\
 \Xhline{2.5\arrayrulewidth}
\end{tabular}
}\vspace{-2mm}
\caption{Topic coherence of different topic models, evaluated on the held-out arXiv data.
\uline{Our full TopicEq model is shown as
``Ours (context + Eq LSTM).''}
}\vspace{0mm}
\label{tbl:coherence_eval}
\end{table}

\begin{table}[t!]
\addtolength{\tabcolsep}{0pt}
\setlength\cellspacetoplimit{3.5pt}
\setlength\cellspacebottomlimit{2.5pt}
\hspace{-2mm}
\scalebox{0.77}{
\begin{tabular}{Sc|Sl}
\Xhline{1.2pt}
	\!\!\scalebox{0.95}{{\bf Quantum physics}} 
	& \scalebox{0.98}[1]{spin energy field electron magnetic state states hamiltonian}\!\!\!\\
\hdashline[2pt/1.5pt]
	\!\scalebox{0.95}{{\bf Particle physics}}& \scalebox{0.98}[1]{higgs neutrino coupling decay scale masses mixing quark}\\
\hdashline[2pt/1.5pt]
	\!\scalebox{0.95}{{\bf Astrophysics}}& \scalebox{0.98}[1]{mass gas star stellar galaxies disk halo radius luminosity}\!\!\!\\
\hdashline[2pt/1.5pt]
	\!\scalebox{0.95}{{\bf Relativity}} & \scalebox{0.98}[1]{black metric hole schwarzschild gravity holes einstein}\!\!\!\!\\
\hdashline[2pt/1.5pt]
	\!\scalebox{0.95}{{\bf Number theory}} & \scalebox{0.98}[1]{prime integer numbers conjecture integers  degree modulo}\\
\hdashline[2pt/1.5pt]
	\!\scalebox{0.95}{{\bf Graph theory}} & \scalebox{0.98}[1]{graph vertex vertices edges node edge number set tree}\\
\hdashline[2pt/1.5pt]
	\!\scalebox{0.95}{{\bf Linear algebra}} & \scalebox{0.98}[1]{matrix matrices vector basis vectors diagonal rank linear\!\!}\\
\hdashline[2pt/1.5pt]
	\!\scalebox{0.95}{{\bf Optimization}} & \scalebox{0.98}[1]{problem optimization algorithm function solution gradient\!\!}\\
\hdashline[2pt/1.5pt]
	\!\scalebox{0.95}{{\bf Probability}} & \scalebox{0.98}[1]{random probability distribution process measure time}\!\!\!\!\\
\hdashline[2pt/1.5pt]
	\!\!\scalebox{0.95}{{\bf Machine learning}} & \scalebox{0.98}[1]{layer word image feature sentence model cnn lstm training}\\
\Xhline{1.2pt}
\end{tabular} }
\vspace{-3mm}
\caption{Topics learned by the TopicEq model. Left: topic name (summarized by us). Right: top words in topic.}
\label{tb:topics}\vspace{-3mm}
\end{table}

\paragraph{Model setting.}
For the inference network  $q(\eta|C)$, we use a 2-layer FFNN
with 300 units, similar to \cite{miao2016neural,miao2017_disc}.
The equation TE-LSTM architecture has two layers and state size 500,
with dropout rate
0.5 applied to each layer \cite{JMLR:v15:srivastava14a}.
The parameters of the TopicEq model are jointly optimized by Adam \cite{kingma2015adam}, with batch size 200, learning rate 0.002, and gradient clipping 1.0 \cite{Pascanu2012}.

\subsection{Topic Model Evaluation}
\label{sec:tm_eval}

We first study the topic modeling performance of TopicEq, by
evaluating the coherence of the learned topics $\beta$ \cite{chang2009reading,newman2010automatic,mimno2011optimizing}.
Specifically, following \cite{lau2014machine}, we compute the normalized PMI metric on the held-out test set.
As our TopicEq model incorporates {joint}, {RNN-based} equation model, to analyze its effect, we compare the full TopicEq model with the following baseline topic models:
\begin{itemize}[topsep=1pt]
    \setlength{\itemsep}{0mm}
    \setlength{\leftskip}{0mm}
    \item LDA (context only): we apply LDA to the word text
    \item Ours (context only): TopicEq without the equation model
    \item Ours (context + Eq BOW): TopicEq's joint LSTM equation model (Eq \ref{eq:lstm_theta}) is replaced by a baseline bag-of-tokens model similar to that for context words.
\end{itemize}
The evaluation results are summarized in Table \ref{tbl:coherence_eval}.
The full TopicEq model is shown as ``Ours (context + Eq LSTM)'' in the table.
We observe that TopicEq's topic model component (2nd row) performs on a par with LDA (1st row),
but it achieves a significant boost (+0.01) when trained together with the LSTM
equation model (4th row).
Adding equations as bag of tokens (3rd row) does improve topic models marginally (+0.002), but the improvement made by using joint LSTM equation model is 5 times greater.
These results show that a joint RNN equation model provides significant information to aid topic modeling of scientific texts.

\begin{table}[t!]
\renewcommand{\arraystretch}{1.05}
\addtolength{\tabcolsep}{0pt}
\hspace{-1.5mm}
\centering
\scalebox{0.9}{
\begin{tabular}{l||cc|c}
\Xhline{2.5\arrayrulewidth}
    \multirow{2}{*}{\hspace{9mm}\textbf{Equation Model} }  &\multicolumn{2}{c|}{\textbf{Perplexity} \vrule width 0pt height 11pt depth 0pt} & \!\textbf{Error \scalebox{0.6}[0.8]{(\%)}}\!\!\!\\
    \multicolumn{1}{c||}{\textbf{} \vrule width 0pt height 9pt depth 4pt} & ~\textbf{50}~ & ~\textbf{100}~ & ~\textbf{100}~\\
     \Xhline{2.5\arrayrulewidth}
 \hspace{-2mm}\scalebox{0.95}{\textbf{No joint training}} \vrule width 0pt height 12pt depth 4.5pt &   &  &  \\
 \!LSTM (no topic)~\vrule width 0pt height 0pt depth 0pt & 5.81 & 5.81   & 15.3 \\
 \!LSTM + LDA~\vrule width 0pt depth 0pt & 5.54 & 5.52  & 13.4 \\
 \hspace{-2mm}\scalebox{0.95}{\textbf{Joint training with topic model}} \vrule width 0pt height 10pt depth 4.5pt &     &  \\
 \!TD-LSTM ~~\,(Lau et al. 2017)\vrule width 0pt height 0pt depth 0pt & 5.44 & 5.41  & 12.5\\
 \!TE-LSTM ~~~(\textbf{Ours}) \vrule width 0pt depth 5pt & {\bf 5.36} & {\bf 5.34}  & {\bf 11.7}\\
 \Xhline{2.5\arrayrulewidth}
\end{tabular}
}\vspace{-2mm}
\caption{Performance of different equation models, evaluated on held-out arXiv data.
We report the perplexity metric (for \# topics 50, 100 if topic info is used), and the syntax error rate of generated \LaTeX\  equations (for \# topics 100).}\vspace{-3mm}
\label{tbl:eq_model_eval}
\end{table}

\paragraph{Why is the RNN helpful?\!\!}

We hypothesize that one
reason why the joint RNN equation model is more helpful than the bag-of-tokens equation model is that the
RNN also captures syntax-level information in equations.
But one might argue that
the introduction of the RNN itself was useful for topic modeling (e.g. as a form of regularization).
To study our hypothesis,
we re-trained TopicEq with each equation's token {order} randomly shuffled in the training data---thus corrupting the syntactic information of each equation.
The result is shown in Table
\ref{tbl:coherence_eval} as ``Ours (context + Eq LSTM shuffled).''
This time, the topic model performance degrades severely and falls to the level of the baseline topic model, ``Ours (context only)''.
This result supports the claim that the original TopicEq's joint RNN actually captured syntactic features of equations, providing more effective help for topic modeling than a bag-of-token equation model does.

This idea also makes intuitive sense. Mathematical equations use a much smaller vocabulary (symbols / variables) than word texts, and thus often need phrase or syntax-level information to aid topic modeling.
For example, in the equations in Figure \ref{fig:context_exm}, phrases like $T_{\mu\nu}$ (use of super/sub-scripts for a tensor) and $\lambda \|x\|_1$ (regularization term) provide rich information to identify the topics (relativity and optimization), while
the corresponding bags of tokens
$\{\mu, \nu, T\}$ and $\{1, \lambda, x, |\}$ themselves do not provide as much help.

\paragraph{Learned topics.}
To
visualize the topic modeling performance, we sampled 10 topics learned by TopicEq
(Table \ref{tb:topics}).
They intuitively reflect the scientific topics of arXiv articles.

\begin{table}[t!]
    \hspace{-0.5mm}
    \includegraphics[width=0.485\textwidth]{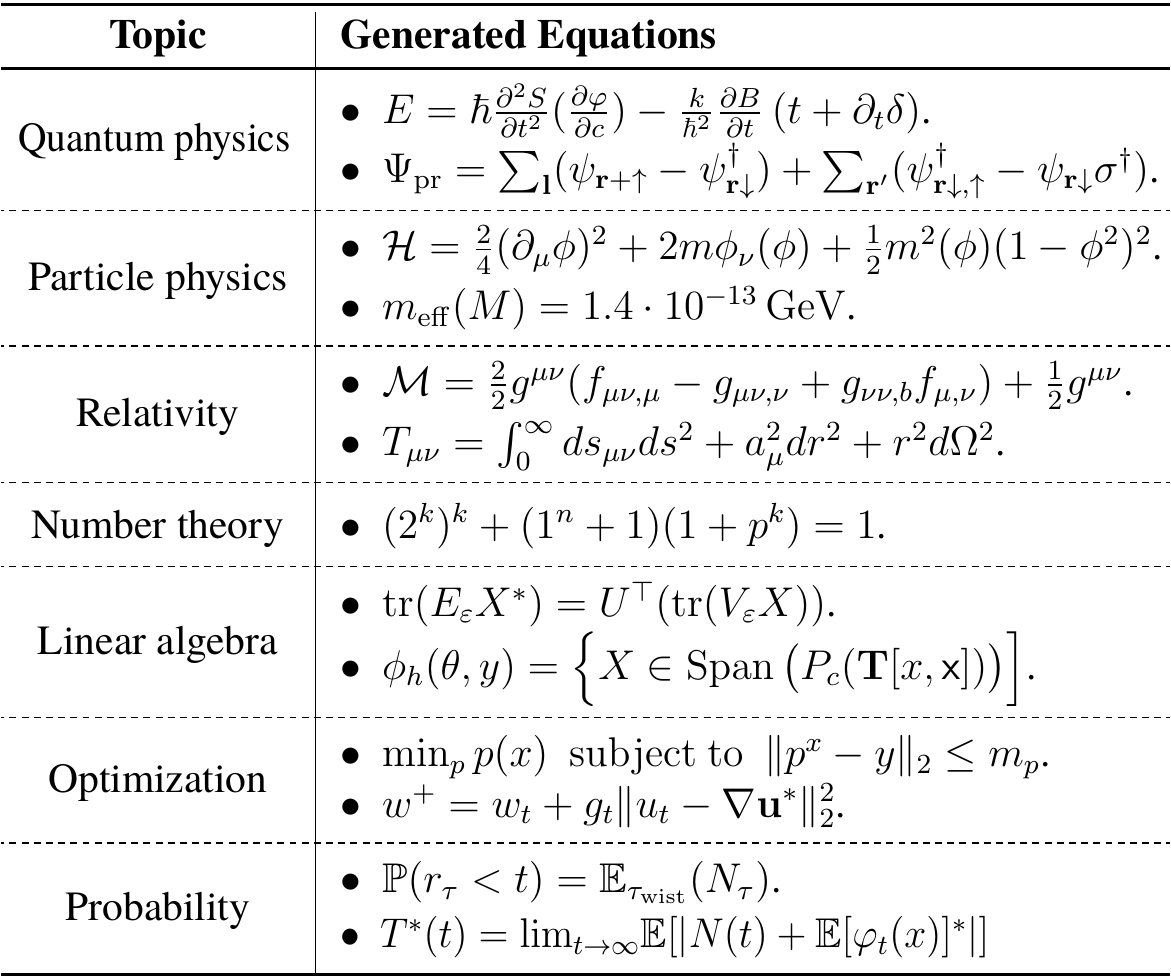}\vspace{-3mm}
    \caption{
    The TopicEq model generates equations that
    reflect the characteristics of given topics. Left: topic (picked from Table \protect\ref{tb:topics}). Right: equations {generated by the model} conditioned on the given topic (one-hot topic vector $\theta$).
    }\vspace{-3mm}
    \label{tbl:gen_eq}
\end{table}

\subsection{Equation Model Evaluation}
\label{sec:eq_model_eval}

Next, we evaluate the equation model component of TopicEq by measuring the test set perplexity.
Additionally, as the grammaticality of equations can be measured using the \LaTeX\ compiler, we also evaluate the syntax error rate of generated equations.
We compare our TE-LSTM with
\begin{itemize}[topsep=1pt]
    \setlength{\itemsep}{0mm}
    \setlength{\leftskip}{2mm}
    \item a generic LSTM (no topic knowledge)
    \item LSTM\,+\,LDA: the topic vector $\theta$ obtained from a pre-trained LDA is concatenated to the output of LSTM
\end{itemize}
and a recent topic-dependent LSTM applied to our task
\begin{itemize}[topsep=1pt]
    \setlength{\itemsep}{0mm}
    \setlength{\leftskip}{2mm}
    \item TD-LSTM \cite{lau2017topically}: $\theta$ is added to the output of LSTM via a dense layer.
\end{itemize}
TD-LSTM and our TE-LSTM are jointly trained with our topic model component.
As Table \ref{tbl:eq_model_eval} shows, all the topic-dependent LSTMs
are superior to the vanilla LSTM in both the perplexity metric and syntax error metric.
Moreover, our TE-LSTM outperforms TD-LSTM, suggesting that the model
better incorporates topic knowledge by embedding $\theta$ inside the LSTM.
We also find that
compared to \cite{wang2017topic}'s Mixture-of-Expert LSTM, our model achieves similar performance in this task while requiring fewer parameters and much less training time (40\% reduction).
In total, compared to the generic LSTM, our TE-LSTM equation model reduces test perplexity by 8\% (relative) and syntax error rate by 3.5\% (absolute).
This result suggests that incorporating context/topic information can improve the quality and grammaticality of equation modeling.

\section{Qualitative Analysis \& Applications}
\label{sec:applications}

\begin{table}[t!]
    \centering
    \includegraphics[width=0.47\textwidth]{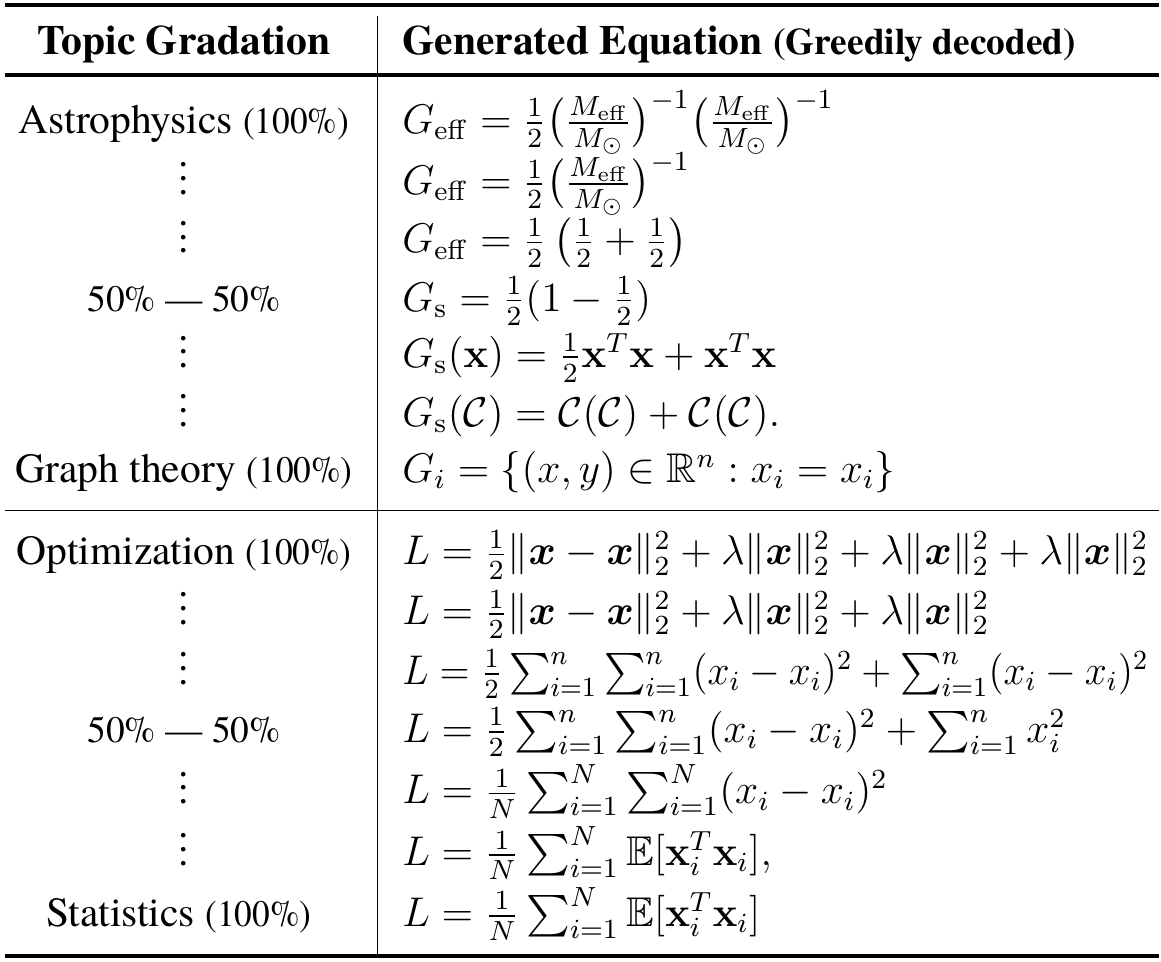}\vspace{-3mm}
    \caption{
    We let the TopicEq model greedily generate equations
    while smoothly changing $\theta$ between two topics (via linear interpolation).
    Left: given topic pair and its interpolation. Right: generated equation (for the first topic pair, we let the model generate from $G$; for the second pair, from $L=$).
    }\vspace{-2mm}
    \label{tbl:gen_eq_gradation}
\end{table}

\begin{table}[t!]
    \hspace{-6mm}
    \includegraphics[width=0.51\textwidth]{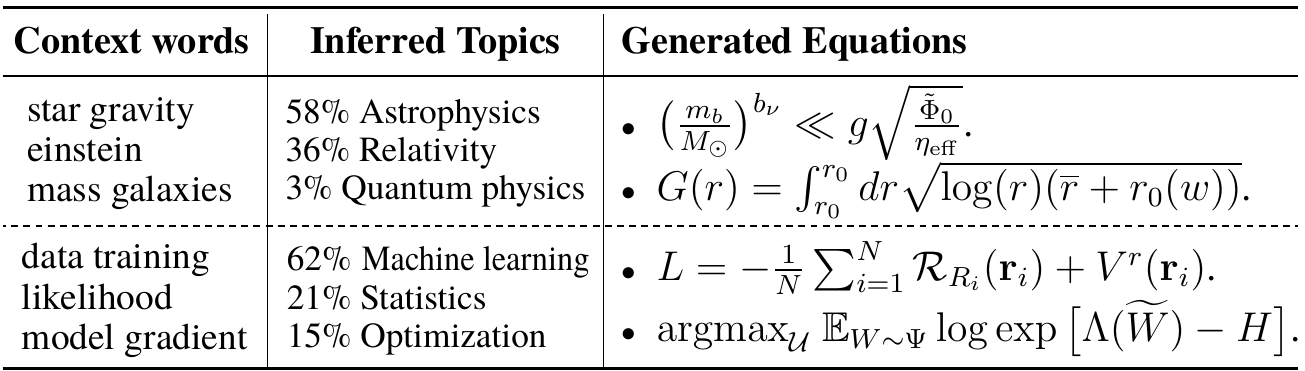}\vspace{-3mm}
    \caption{
    Given a set of context words picked from an article abstract (1st column), we let TopicEq {infer its topic proportions} (2nd col) and {generate equations} (3rd col).
    }\vspace{-3mm}
    \label{tbl:gen_eq_mix}
\end{table}

\begin{table*}[!ht]
\definecolor{mydarkgreen}{HTML}{008000}
\definecolor{mydarkred}{HTML}{dc143c}
\definecolor{mydarkyellow}{HTML}{ff8c00}
\newcommand\bunsuu[2]{
\dfrac{\,\lower.44ex\hbox{$#1$}\,}{\,\lower-.1ex\hbox{$#2$}\,} }

\centering
\renewcommand\cellalign{lc}
\setlength\cellspacetoplimit{2pt}
\setlength\cellspacebottomlimit{2pt}
\addtolength{\tabcolsep}{0pt}
\scalebox{0.96}{
\begin{tabular}{Sl || Sl | Sl} \Xhline{2\arrayrulewidth}
\multicolumn{1}{Sc ||}{
\multirow{2}{*}{
\makecell{\vrule width 0pt height 11pt depth 5pt
\hspace{15mm}\scalebox{0.9}{\textbf{Given Equation}}\\[-1mm]
\scalebox{0.8}{[\![~]\!] shows the correct formula name for readers}
} }} &
\multicolumn{2}{Sc}{\vrule width 0pt height 9pt depth 3pt
\scalebox{0.9}{\textbf{Inferred Topic}}
~\,\scalebox{0.8}{\textbf{(showing top 5 words)}}
}\!
\\ \cdashline{2-3}[2pt/1.5pt]
&  \multicolumn{1}{c|}{\vrule width 0pt height 11pt depth 4pt \scalebox{0.85}{\textbf{by our TopicEq}} } & \multicolumn{1}{c}{\scalebox{0.84}[0.85]{\textbf{by bag-of-token baseline}}} \\
\Xhline{2\arrayrulewidth}
 \makecell{
    \!\scalebox{0.9}{\textbf{\#1}} ~
    \scalebox{0.9}{$\displaystyle i \hbar \scalebox{0.8}{$\bunsuu{\partial}{\partial t}$}
    \vert\Psi(\mathbf{r},t)\rangle = \hat H\vert\Psi(\mathbf{r},t)\rangle $} ~
    \scalebox{0.55}{[\![Schr\"{o}dinger Equation]\!]} 
    } &
    \makecell{
    \scalebox{0.8}[0.82]{hamiltonian, spin, particle,}\\[-1.3mm]
    \scalebox{0.8}[0.82]{interaction, wave}~~\raisebox{-2pt}{\textcolor{mydarkgreen}{\ding{51}}}
    } &
    \makecell{
    \scalebox{0.8}[0.82]{time, operator, space, }\\[-1.3mm]
    \scalebox{0.8}[0.82]{hamiltonian, system}~~\raisebox{-2pt}{\textcolor{mydarkgreen}{\ding{51}}}
    }
 \\ \hdashline[2pt/1.5pt]
 \makecell{
    \!\scalebox{0.9}{\textbf{\#2}} ~ \scalebox{0.9}{$F = \scalebox{0.9}{$\bunsuu{d(mv)}{dt}$} $} ~
    \scalebox{0.55}{[\![Newton's 2nd Law of Motion]\!]}
    } &
    \makecell{
    \scalebox{0.8}[0.82]{velocity, particle, pressure,}\\[-1.3mm]
    \scalebox{0.8}[0.82]{motion, force}~~\raisebox{-2pt}{\textcolor{mydarkgreen}{\ding{51}}}
    }
    &
    \makecell{
    \scalebox{0.8}[0.82]{time, velocity, particle,}\\[-1.3mm]
    \scalebox{0.8}[0.82]{diffusion, force}~~\raisebox{-2pt}{\textcolor{mydarkgreen}{\ding{51}}}
    }
\\ \hdashline[2pt/1.5pt]
 \makecell{\vrule width 0pt depth 6pt height 12pt
    \!\scalebox{0.9}{\textbf{\#3}} ~
    \scalebox{0.9}{$ W + \Delta U =$ \scalebox{0.8}{$\displaystyle\int$} $f \cdot { d } x - m g h $} ~
    \scalebox{0.55}{[\![Potential energy \& Work]\!]}\vspace{-1mm}
    } &
    \makecell{
    \scalebox{0.8}[0.82]{direction, force, surface,}\\[-1.3mm] \scalebox{0.8}[0.82]{strain, stress}~~~~\raisebox{-2pt}{
    \scalebox{1.5}[1]{\textcolor{mydarkyellow}{\textbf{\textit{?}}}}
    }
    }
    &
    \makecell{
    \scalebox{0.8}[0.82]{method, order, solution, }\\[-1.3mm] \scalebox{0.8}[0.82]{numerical, problem}~~\raisebox{-2pt}{\scalebox{1.1}{\textcolor{mydarkred}{\ding{55}}}}
    \!\!\!
    \raisebox{-1pt}{\textcolor{mydarkred}{\scalebox{0.75}[0.8]{{ \sf \sl (vague)}}}}\!\!\!\!
    }
\\ \hdashline[2pt/1.5pt]
 \makecell{\vrule width 0pt depth 6pt height 12pt
    \!\scalebox{0.9}{\textbf{\#4}} ~
    \scalebox{0.9}{$f_m = \sigma ( W_f h_{m-1} + U_f x_m + b_f )$} ~
    \scalebox{0.55}{[\![LSTM]\!]}\vspace{-0.2mm}
    } &
    \makecell{
    \scalebox{0.8}[0.82]{layer, word, image,}\\[-1.3mm]
    \scalebox{0.8}[0.82]{feature, network}~~\raisebox{-2pt}{\textcolor{mydarkgreen}{\ding{51}}}
    }
    &
    \makecell{
    \scalebox{0.8}[0.82]{function, section, problem,}\\[-1.3mm] \scalebox{0.8}[0.82]{condition, solution}~~\raisebox{-2pt}{\scalebox{1.1}{\textcolor{mydarkred}{\ding{55}}}}
    \!\!\!
    \raisebox{-1pt}{\textcolor{mydarkred}{\scalebox{0.75}[0.8]{{ \sf \sl (vague)}}}}\!\!\!\!
    }

\\ \hdashline[2pt/1.5pt]
 \makecell{
    \!\scalebox{0.9}{\textbf{\#5}} ~ \scalebox{0.9}{$\displaystyle P(X|Y) = \scalebox{0.9}{$\bunsuu{P(Y|X)P(X)} {P(Y)}$} $} ~
    \scalebox{0.55}{[\![Bayes' Theorem]\!]}\vspace{-0.2mm}
    } &
    \makecell{
    \scalebox{0.8}[0.82]{random, variable, probability,}\\[-1.3mm]
    \scalebox{0.8}[0.82]{distribution, entropy}~~\raisebox{-2pt}{\textcolor{mydarkgreen}{\ding{51}}}\!\!\!
    }
    &
    \makecell{
    \scalebox{0.8}[0.82]{probability, random, theorem\!\!}\\[-1.3mm]
    \scalebox{0.8}[0.82]{variable, distribution}~~\raisebox{-2pt}{\scalebox{1}{\textcolor{mydarkgreen}{\ding{51}}}}\!\!\!
    }
\\ \hdashline[2pt/1.5pt]
 \makecell{
    \!\scalebox{0.9}{\textbf{\#6}} ~ \scalebox{0.88}[0.9]{$ {\textstyle \lim_{n\to\infty}} P\left(\sqrt{n}(S_n-\mu) \le z\right) = \Phi\left(\frac{z}{\sigma}\right)$}~ \\[-1.2mm] ~~~~
    \scalebox{0.55}{[\![Central Limit Theorem]\!]}\!\!\vspace{-0.5mm} \vspace{-0.5mm}
    } &
    \makecell{
    \scalebox{0.8}[0.82]{measure, random, process,}\\[-1.3mm]
    \scalebox{0.8}[0.82]{gaussian, convergence}~~\raisebox{-2pt}{\textcolor{mydarkgreen}{\ding{51}}}\!\!\!
    }
    &
    \makecell{
    \scalebox{0.8}[0.82]{probability, random, theorem\!\!}\\[-1.3mm]
    \scalebox{0.8}[0.82]{variable, distribution}~~\raisebox{-0pt}{\scalebox{0.6}[0.6]{\textcolor{mydarkyellow}{\ding{51}}}}\!\!\!
    }
\\ \hdashline[2pt/1.5pt]
 \makecell{
    \!\scalebox{0.9}{\textbf{\#7}} ~ \scalebox{0.9}[0.9]{$f(x) = f(a)+\frac {f'(a)}{1!} (x\!-\!a)+ \frac{f''(a)}{2!} (x\!-\!a)^2+ \cdots $} \\[-1.2mm] ~~~~
    \scalebox{0.55}{[\![Taylor Expansion]\!]}\vspace{-1mm}
    } &
    \makecell{
    \scalebox{0.8}[0.82]{coefficients, series, expansion}\\[-1.3mm]
    \scalebox{0.8}[0.82]{fourier, polynomial}~~\raisebox{-2pt}{\textcolor{mydarkgreen}{\ding{51}}}\!\!\!
    }
    &
    \makecell{
    \scalebox{0.8}[0.82]{polynomial, series, function,}\\[-1.3mm]
    \scalebox{0.8}[0.82]{convergence, order}~~\raisebox{-2pt}{\textcolor{mydarkgreen}{\ding{51}}}\!\!\!
    }\!\!\!
\\ \hdashline[2pt/1.5pt]
 \makecell{
    \!\scalebox{0.9}{\textbf{\#7}~\!\scalebox{1.5}[1]{$\boldsymbol{'}$}} \, \scalebox{0.9}[0.9]{$h(b) = h(a)+\frac {h'(b)}{1!} (b\!-\!a)+ \frac{h''(b)}{2!} (b\!-\!a)^2+ \cdots $} \\[-1.2mm] ~~~~
    \scalebox{0.55}{[\![Taylor Expansion]\!]}\vspace{-1mm}
    } &
    \makecell{
    \scalebox{0.8}[0.82]{coefficients, series, expansion}\\[-1.3mm]
    \scalebox{0.8}[0.82]{fourier, polynomial}~~\raisebox{-2pt}{\textcolor{mydarkgreen}{\ding{51}}}\!\!\!
    }
    &
    \makecell{
    \scalebox{0.8}[0.82]{function, integral, equation}\\[-1.3mm]
    \scalebox{0.8}[0.82]{point, solution}~~\raisebox{-2pt}{\scalebox{1.1}{\textcolor{mydarkred}{\ding{55}}}
    }\!\!\!
    \raisebox{-1pt}{\textcolor{mydarkred}{\scalebox{0.8}{{ \sf \sl (fooled)}}}}\!\!\!
    }\!\!\!
\\\Xhline{2\arrayrulewidth}
\end{tabular}}\vspace{-1.5mm}
    \caption{The TopicEq model can infer the appropriate topic for equations from various domains, with better precision and consistency than bag-of-token baseline. Left: given equation. Right: topic {inferred by our model} and the baseline.
    \textcolor{mydarkgreen}{\ding{51}} indicates that the inferred topic is correct; \textcolor{mydarkred}{\ding{55}} not good.
    We verified that the {\it exact} same equations did not appear in the training data.
    }\vspace{-3mm}
    \label{tbl:eq_cap}
\end{table*}

\subsection{Topic-aware Equation Generation}
The TopicEq model can
generate meaningful equations from specified topics, using Eq \ref{eq:lstm_theta} (TE-LSTM).
For example, given a topic $k$, we let $\theta$ be the one-hot vector representing the topic;
conditioned on $\theta$, and
starting from 
\scalebox{0.7}[0.9]{\texttt{\textbf{<START>}}} token, we keep sampling the next \LaTeX\ token until the \scalebox{0.7}[0.9]{\texttt{\textbf{<END>}}} token is generated.
Table \ref{tbl:gen_eq}
shows
several topics picked from Table \ref{tb:topics} (left), and
equations generated from each of these topics (right).
We see that the artificial equations generated by the model
clearly reflect the distinctive characteristics of the given topics.
For instance, derivatives, and number\,+\,units are generally used for physics; electron configuration \scalebox{0.8}{$\uparrow, \downarrow$} for quantum physics;
series of tensors like $T_{\mu\nu}$ for relativity;
prime number $p$ for number theory;
$\mathbb{E}$, $\mathbb{P}$ clauses for probability.
We also note that the equations generated by our TE-LSTM use not only topic-specific symbols but also topic-specific phrases and syntax (e.g., a set definition is used for linear algebra; ``$\min$ subject to'' clause for optimization).
These qualitative results support that TopicEq is capable of fully incorporating topic information for equation modeling.

\paragraph{Mixtures of topics.}
The model can also generate equations from a mixture of topics by setting $\theta$ accordingly.
To qualitatively analyze the space of the topic vector $\theta$ in terms of equation generation,
we let the model generate equations while smoothly changing $\theta$ between two topics (i.e., one-hot vectors $\theta_1$ and $\theta_2$) via linear interpolation: $\theta (t) = (1\!-\!t)\theta_1 + t \theta_2$ for $t\!\in\! [0,1]$.
In Table \ref{tbl:gen_eq_gradation}, for two examples we
show the given topic pair and its interpolation (left), and the equation greedily decoded from each $\theta (t)$ (right).
We let the model start all equations from $G$ in the first example (astrophysics and graph theory), and from $L=$ in the second example (optmization and statistics).
In both cases
we observe that the generated equations make a smooth transition from one topic to the other --- e.g., for the first example,
from using $M_{\mathrm{eff}} /M_{\odot} $ (astrophysics) to
using linear algebraic term ${\bf x}^T {\bf x}$,
and finally a set notation (graph theory).
In the second example, where the two topics optimization and statistics are closely related, the generated equations make a very intuitive transition:
from an optimization objective with norms and regularization terms (top),
to using summation terms (middle) and finally expectations (bottom; statistics topic).
These observations support that {TopicEq  learns smooth representations for the latent topic vector $\theta$} (especially for a mixture of closely related topics), regarding equation generation.

Finally, we illustrate that the model can generate  equations from a given set of context words.
Specifically, we let the model infer the topic proportion $\theta$ of the context words via the inference network $q(\eta\hspace{0.5pt}|\hspace{0.5pt}C)$, and then generate equations from $\theta$ via Eq \ref{eq:lstm_theta} (TE-LSTM).
As Table \ref{tbl:gen_eq_mix} shows,
the model is able to infer the right topic mixture (2nd column)
and generate equations
that reflect those topics (e.g., solar mass $M_{\odot}$ and radius $r$ are used for the top example; loss function $L$, $\arg \max$, and $\mathbb{E}$ for the bottom example).

\subsection{Equation Topic Inference}

Identifying the topic of equations is an important task that allows readers to obtain semantic descriptions for equations unfamiliar to them.
However, while some work \cite{schubotz2016semantification,stathopoulos2018variable} has studied the task of identifying the meaning of individual mathematical symbols, no prior work has succeeded in providing descriptions to entire equations from various domains.

Our TopicEq model can be utilized
to identify the topic of given equations. Specifically,
with a trained TopicEq model,
for a given equation $\eq$, we find the topic $k \in [K]$
(so $\theta$ is a one-hot vector) that maximizes the likelihood
$p(eq \,|\,\theta)$ in Eq \ref{eq:lstm_theta}, which is parametrized by our topic-dependent LSTM.
Table \ref{tbl:eq_cap} shows examples of equations across different domains (1st column), and the most likely topic inferred by our model for each equation (2nd column). We used $K\!=\!100$ topics in this task.
We observe that the TopicEq model correctly identifies the domains or even finer topics
(e.g., note the distinction between \textbf{\#5} and \textbf{\#6}) for most of the
given equations.

\begin{table*}[h]
\centering
\renewcommand\cellalign{lc}
\setlength\cellspacetoplimit{1pt}
\setlength\cellspacebottomlimit{1pt}
\addtolength{\tabcolsep}{1pt}
\scalebox{0.78}{
\begin{tabular}{Sc || Sc || Sc | Sc | Sc} \Xhline{3\arrayrulewidth}
\multirow{2}{*}{
\makecell{ \vrule width 0pt height 12pt depth 0pt
\textbf{Math}\\[-0.7mm] \!\!\textbf{symbol}\!}} &  \multicolumn{4}{c}{\vrule width 0pt height 11pt depth 5pt \textbf{Topics}} \\\cdashline{2-5}[2pt/1.5pt]
 & \vrule width 0pt height 10pt depth 3pt \textbf{{No Topic}} & \textbf{{Probability}} & \textbf{{Quntum physics}} & \textbf{{Graph theory}}\\
\Xhline{3\arrayrulewidth}
 \makecell{\vrule width 0pt height 12pt depth 0pt \scalebox{1.15}{$E$}}
 &
 \makecell{energy, expectation, elliptic curve}
 &
 \makecell{expectation, expected value}
 &
 \makecell{electric field, energy}
 &
 \makecell{edge, \scalebox{0.95}[1]{spectral sequence}}
 \\ \hdashline[2pt/1.5pt]
 \makecell{\vrule width 0pt height 12pt depth 0pt \scalebox{1.15}{$M$}}
 &
 \makecell{mass, matrix}
 &
 \makecell{martingale, maximum}
 &
 \makecell{magnetic moment, mass}
 &
 \makecell{matroid, matching}
 \\ \hdashline[2pt/1.5pt]
 \makecell{\vrule width 0pt height 12pt depth 0pt \scalebox{1.15}{$p$}}
 &
 \makecell{polynomials, momentum, probability}
 &
 \makecell{probability, poisson, distribution}
 &
 \makecell{momentum, proton, pressure}
 &
 \makecell{path, perimeter, probability}
 \\ \hdashline[2pt/1.5pt]
 \makecell{\vrule width 0pt height 12pt depth 0pt \scalebox{1.15}{$T$}}
 &
 \makecell{temperature, transpose,\\[-0.7mm] transfer matrix}
 &
 \makecell{stopping time, test statistic}
 &
 \makecell{temperature,\\[-0.7mm] thermal conductivity}
 &
 \makecell{tree, trees,\\[-0.7mm] triangulation}
 \\ \hdashline[2pt/1.5pt]
 \makecell{\vrule width 0pt height 12pt depth 0pt \scalebox{1.15}{$V$}}
 &
 \makecell{potential, voltage, visibility, volume}
 &
 \makecell{variance, volatility}
 &
 \makecell{voltage, potential energy}
 &
 \makecell{vertex, volume, \scalebox{1}[1]{SVD}}
 \\ \hdashline[2pt/1.5pt]
 \makecell{$\sigma$}
 &
 \makecell{conductivity, variance,\\[-0.7mm] normal distribution}
 &
 \makecell{standard deviation,\\[-0.7mm] normal distribution}
 &
 \makecell{conductivity,\\[-0.7mm] pauli matrices}
 &
 \makecell{permutation, simplex}
 \\ \hdashline[2pt/1.5pt]
 \makecell{\vrule width 0pt height 11pt depth 0pt \scalebox{1.15}{$\mid$}}
 &
 \makecell{norm, distance, conditional}
 &
 \makecell{conditional probability}
 &
 \makecell{absolute value}
 &
 \makecell{triangle inequality, cardinality}
 \\
\Xhline{3\arrayrulewidth}
\end{tabular}}\vspace{-2mm}
\caption{Top word phrases \textbf{predicted by our topic-aware alignment model} for each math symbol. We show the prediction results for three of the learned topics (3rd-5th column), as well as the non-topic baseline (2nd column).}\vspace{-3mm}
\label{tbl:alignment_demo}
\end{table*}

\paragraph{Is an RNN necessary for this task?}
We repeated this experiment using a bag of tokens model for equations in Eq \ref{eq:lstm_theta}  (instead of LSTM),
to analyze whether the RNN equation model
provides an advantage over the bag of tokens-based approach in this task.
As can be seen in Table \ref{tbl:eq_cap}, 3rd column,
this bag-of-tokens baseline performs as well in
\textbf{\#1} and \textbf{\#2}, which have topic-specific variables like $\hbar$, $\psi$, $v$, but fails in \textbf{\#3} and \textbf{\#4}, which consist of a relatively generic set of symbols $\{f, h, m, U, W, x\}$ and require recognizing phrases like $\int f \cdot dx$ (work)
and $ \sigma(Wh+b)$ (neural network layer) to identify the correct topic.
Indeed, the topics predicted for \textbf{\#3} and \textbf{\#4} are very generic and similar.
Similarly,
the bag-of-tokens baseline fails to distinguish \textbf{\#5} and \textbf{\#6}, most likely because it does not recognize the phase and syntax-level differences between these two equations.
Finally, for \textbf{\#7} (Taylor Expansion), we also experimented with  \textbf{\#7'}, where we just changed some variable names without altering the equation's meaning and syntax.
While our TopicEq still recognizes this to be the same topic as \textbf{\#7},
the bag-of-tokens baseline is fooled by the changed variable names and predicts a wrong topic.
These observations suggest that the RNN equation model can capture phrase and syntax-level information, and can consistently infer the correct topics for equations from various domains.
The TopicEq model could be used to help readers interpret equations unfamiliar to them.

\section{Extension: Topic-aware alignment between mathematical tokens and words}

\noindent
Mathematical symbols (including variables) carry different meanings in different contexts or topics.
Prior work \cite{pagael2014mathematical,schubotz2016semantification,stathopoulos2018variable}
has studied the task of identifying meanings of math variables using surrounding words, but
its topic dependence has not been modeled explicitly.
Here we present a variant of the TopicEq model that captures \textit{topic-dependent} alignment between mathematical tokens and words from scientific document data. Specifically, we aim to learn the most probable descriptions (word phrases) $w$ associated with a given math symbol $s$, under a given topic or topic mixture $\theta$: $p(w \,|\, s, \theta)$.

\paragraph{Baseline alignment model.}
We use the equations and context texts from our \textit{ContextEq} corpus.
Similar to \cite{pagael2014mathematical}, we consider that the descriptions of math symbols often appear in the sentence immediately before or after the given equation (\textit{immediate context}).
We then consider a simple alignment model
between symbols $s$ in the equation and phrases $w$ in the immediate context, such that
\begin{align}
    w \sim \text{Mult}(\text{softmax}(A \boldsymbol{s}))
\end{align}
Here vector $\boldsymbol{s}\!\in\!\mathbb{R}^L $ is the bag-of-tokens representation of the equation. $A\!\in\!\mathbb{R}^{M\times L} $ is the alignment matrix we estimate from the data, by maximizing the likelihood $p(w\,|\,\boldsymbol{s})$.
$L, M$ are the vocab sizes of symbols and word descriptions.
For the vocabulary of word descriptions,
we collect the titles of Wikipedia pages that contain mathematical equations.
We then use the top 2,000 phrases that appear in our arXiv dataset.
For math symbols, we use the top $L\!=\!200$.

To predict
$w$ given a single symbol $s$, we set $\boldsymbol{s}$ to be the one-hot vector representing $s$, as a surrogate.

\paragraph{Topic-aware alignment model.}
To model $p(w \,|\, s, \theta)$, we want the alignment matrix $A$ to depend on $\theta$. Motivated by the tensor factorization method in \cite{song2016factored},
we let
\begin{align}
    A(\theta) = W_a \cdot \text{diag} (W_b \theta) \cdot W_c
\end{align}
where $W_a \!\in\! \mathbb{R}^{M\times F}$, $W_b \!\in\! \mathbb{R}^{F\times K}$, $W_c \!\in\! \mathbb{R}^{F\times L}$
are parameters to estimate. $F$ is the number of factors, which we set to be equal to the number of topics $K$.
To jointly perform topic modeling and alignment learning, we consider a variant of TopicEq, where we just replace Eq \ref{eq:lstm_theta} by this topic-dependent alignment model. We train it on the \textit{ContextEq} corpus.

\subsection{Results and Discussion}

Table \ref{tbl:alignmnet_eval} shows the perplexity of the baseline \!/\! topic-aware alignment models evaluated on the held-out test set.
We observe that the
topic information significantly
improves the alignment between math symbols and word descriptions, reducing the perplexity by more than 33\% (relative).

\begin{table}[t!]
\renewcommand{\arraystretch}{1.05}
\addtolength{\tabcolsep}{0pt}
\hspace{-2mm}
\centering
\scalebox{0.85}{
\begin{tabular}{l||cl}
\Xhline{2.5\arrayrulewidth}
     \multicolumn{1}{c||}{\textbf{Alignment Model} \vrule width 0pt height 11pt depth 5pt} & ~\textbf{50}~ & \textbf{100}~  \scalebox{0.7}[0.75]{(\# Topics)\!\!\!\!}\\\Xhline{2.5\arrayrulewidth}
 Baseline (no topic) \vrule width 0pt height 11pt depth 0pt &  602 & 602 \\
 Topic-Aware  \vrule width 0pt depth 4.5pt &  \textbf{406} &  \textbf{387} \\
 \Xhline{2.5\arrayrulewidth}
\end{tabular}
}\vspace{-2mm}
\caption{Test perplexity for phrase prediction.
}\vspace{-4mm}
\label{tbl:alignmnet_eval}
\end{table}

\begin{table}[t!]
\renewcommand{\arraystretch}{1.05}
\addtolength{\tabcolsep}{0pt}
\centering
\scalebox{0.85}{
\begin{tabular}{l||cl}
\Xhline{2.5\arrayrulewidth}
     \multicolumn{1}{c||}{\textbf{Topic Model} \vrule width 0pt height 11pt depth 5pt} & ~\textbf{50}~ &  ~\textbf{100}~  \scalebox{0.7}[0.75]{(\# Topics)\!\!\!\!}\\\Xhline{2.5\arrayrulewidth}

 \!Context Only \vrule width 0pt height 11pt depth 0pt & .085 & .084  \\
 \!with joint \scalebox{1}[1]{Alignment Model}~\vrule width 0pt depth 4.5pt & \textbf{.088}   &  \textbf{.087} \\
 \Xhline{2.5\arrayrulewidth}
\end{tabular}
}\vspace{-2mm}
\caption{Topic coherence evaluation for each topic model.}\vspace{-4mm}
\label{tbl:topic_alignmnet_eval}
\end{table}

~\vspace{-3mm}

\noindent
\textbf{Qualitative results.~~}
Table \ref{tbl:alignment_demo} shows the actual top phrases predicted by the alignment models for several math symbols that are used in a wide range of domains.
The proposed TopicEq variant indeed learns the topic-dependent alignment between symbols and words. For instance,
it associates $E$ with ``expectation'' for the probability topic, ``electric field'' for quantum physics, and ``edge'' for graph theory, which makes intuitive sense.
On the other hand,
the baseline (no topic) model associates $E$ with ``energy'', which is simply the description that appears most frequently across all articles.\\
This is another example where the TopicEq framework can be used to capture the relation of topics and mathematics.

\paragraph{Utility.}
We also note that our topic-aware alignment model can be conditioned on a mixture of topics by setting $\theta$ accordingly.
Given a context text and equation, this model can infer the topic proportion by the topic model component, and then use the topic-aware alignment component to infer
the most probable meaning of each variable in the given equation.
This could aid readers to comprehend scientific documents containing mathematics unfamiliar to them.

\paragraph{Effect on topic modeling.}
In Table \ref{tbl:topic_alignmnet_eval}, we compare  our baseline topic model (top) and this TopicEq variant with the alignment component (bottom).
The joint alignment model provides moderate improvements for topic modeling quality.

\section{Conclusion}
Motivated by the topical correspondence between text and mathematical equations observed in scientific documents,
we proposed \textit{TopicEq}, a joint topic-equation model that generates the text by a topic model and the equations by a topic-dependent RNN.
This joint model outperforms
existing topic models and equation models for scientific texts.
We also qualitatively analyzed TopicEq, and showed its applications and extensions, such as
equation topic inference and topic-aware alignment of mathematical symbols and words.

\subsection*{Acknowledgments}
We thank Matt Bonakdarpour, Paul Ginsparg, Samuel Helms, and
Kriste Krstovski for their assistance, and Jungo Kasai as well as the anonymous reviewers for their feedback.
This work was supported in part by a grant from the Alfred P. Sloan Foundation.

\bibliography{aaai19}
\bibliographystyle{aaai}

\end{document}